\documentclass{PoS}

\title{SW Sex Stars Then and Now: A Review}

\ShortTitle{SW Sex Stars Then and Now: A Review}

\author{\speaker{Linda Schmidtobreick}\\
         European Southern Observatory, Casilla 19001, Santiago 19, Chile\\
         E-mail: \email{lschmidt@eso.org}}

\abstract{SW Sextantis stars are a class of cataclysmic variables originally 
defined via certain peculiar properties that they all have in common. 
In this article, I review
our knowledge of these stars and show
the way from a phenomenological classification to a physical understanding
of these systems. The fact that
SW\,Sex stars accumulate at the upper edge of the period gap is discussed
with respect to the secular evolution of cataclysmic variables.
         }

\FullConference{The Golden Age of Cataclysmic Variables and Related Objects - III, Golden2015 \\
		7-12 September 2015\\
		Palermo, Italy}

\begin{document}

\section{Some History}
\begin{figure}[b]
\resizebox{0.48\hsize}{!}{
\includegraphics[clip=true]{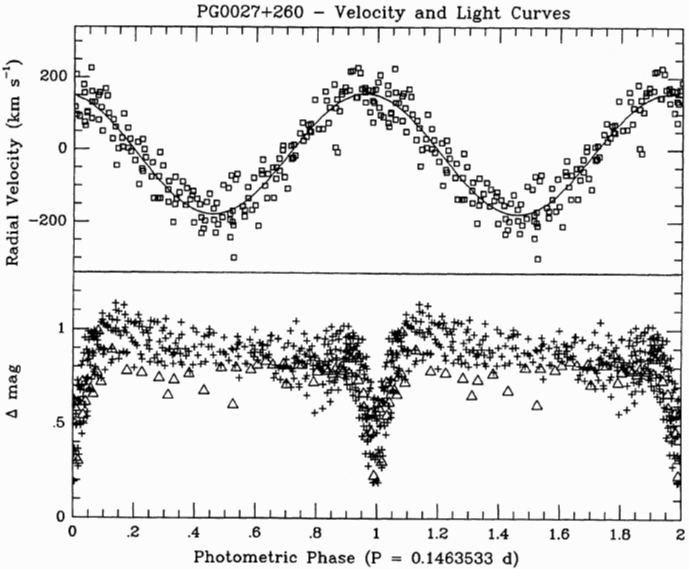}}
\resizebox{0.52\hsize}{!}{
\includegraphics[clip=true]{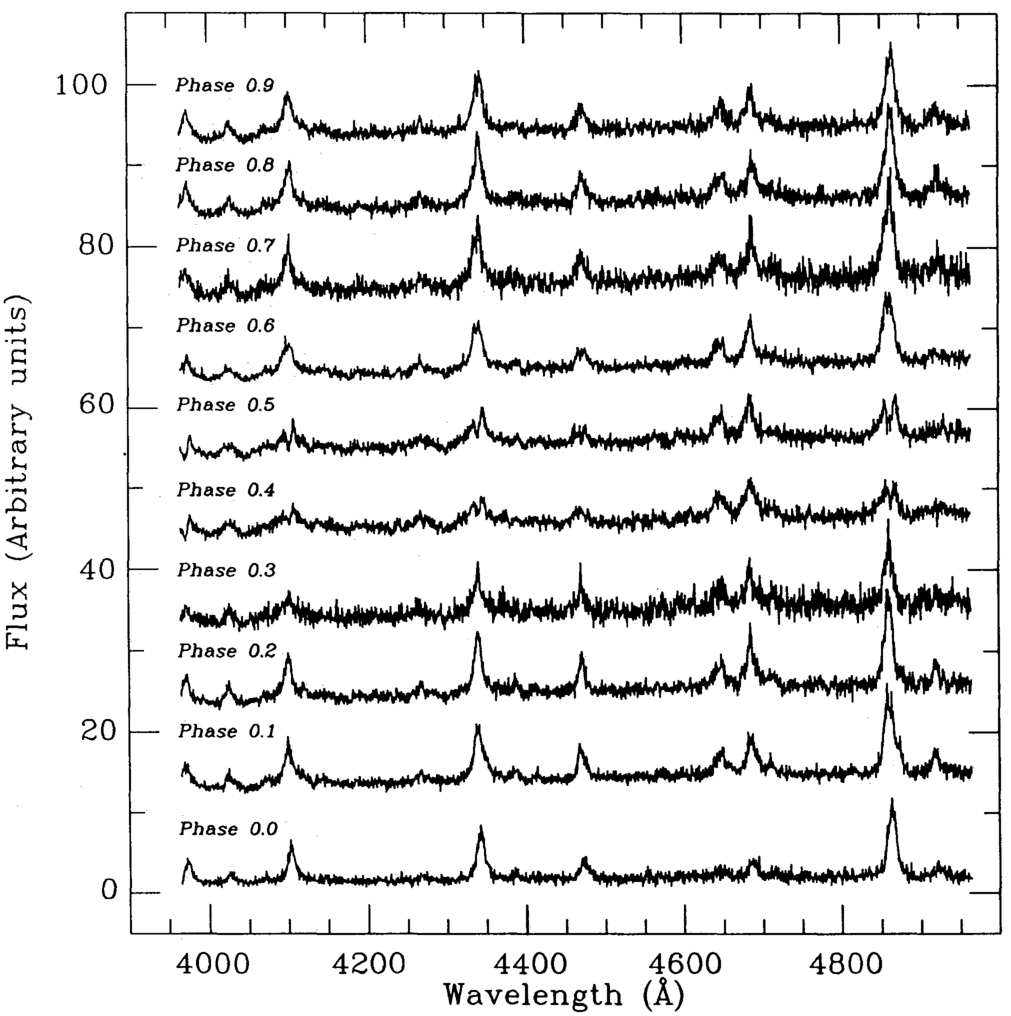}}
\caption{\footnotesize Some examples for the originally defining features of
SW\,Se stars. On the left side, the radial velocities of H$\alpha$ (top) and
differential V magnitudes (bottom) of PX\,And are plotted against the 
orbital phase. The zero-crossing towards negative velocities, which should occur
at phase 0 and 1 for the motion of the white dwarf, happens around phase 
1.2; the velocities just lag behind the expected ones 
\cite{thorstensenetal91-1}).  
On the right side, the phase-resolved spectra of V1315\,Aql show the 
absorption feature at phase 0.5, the eclipse is at 0.0 \cite{dhillon90-1}.
}
\label{fig_V1315Aql_spec}
\end{figure}

About 25 years ago, the mysterious but consistent and reproducible
behaviour of four stars was 
mentioned for the first time by Vik Dhillon \cite{dhillon90-1} and 
Thorstensen et al.\ \cite{thorstensenetal91-1}. 
The stars, SW\,Sex, DW\,UMa, V1315\,Aql and PX\,And were characterised as 
eclipsing novalike systems which however showed single-peaked emission 
lines (see e.g. Figure \ref{fig_V1315Aql_spec}) instead of the
double-peaked lines one would expect for high-inclination accretion discs \cite{horne+marsh86-1}.
The orbital periods of these systems lie in the range between 3 and 4 hours, 
i.e. the region above the period gap. 
They seemed to be non-magnetic, 
as no polarisation or X-rays had been found. 
Nevertheless, He\'II was noted as strong, about half the strength 
of H$\beta$ which is usually a sign for magnetic CVs. 
Thorstensen et al. point out that "the radial velocities of the emission
lines vary periodically, but the Balmer line velocities lag
substantially behind the phase one expects for the white
dwarf on the basis of the eclipse."\cite{thorstensenetal91-1} (see Figure \ref{fig_V1315Aql_spec}
for an example). The final defining feature are 
transient absorption lines that appear only during opposite eclipse phases; 
an example is shown in Figure \ref{fig_V1315Aql_spec}.
The group of stars that fullfill these constraints is referred to as the
SW\,Sex stars (see Table \ref{tab_swsex} for a summary).

\begin{figure}[p]
\centerline{
\resizebox{!}{10.8cm}{\includegraphics{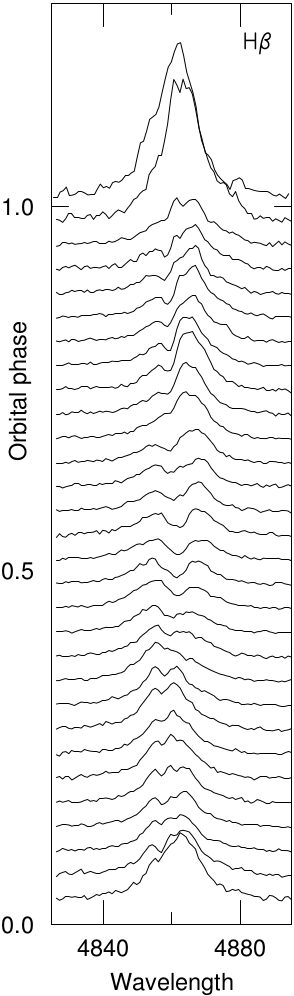}}\hspace{0.5cm}
\resizebox{!}{10.9cm}{\includegraphics{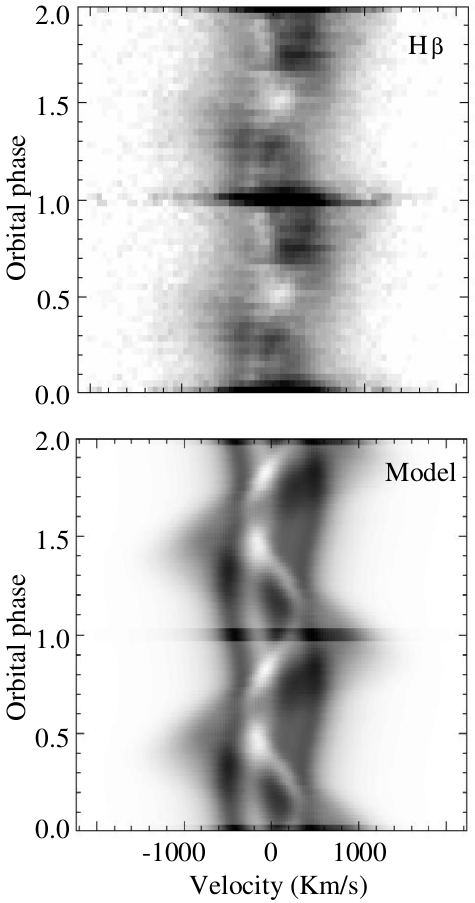}}\hspace{0.5cm}
\begin{minipage}[b]{4.9cm}
\resizebox{4.9cm}{!}{\includegraphics{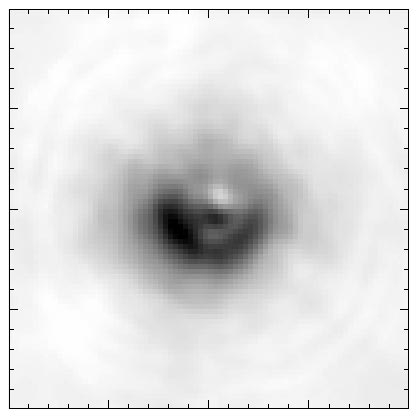}}\\[0.3cm]
\resizebox{4.9cm}{!}{\includegraphics{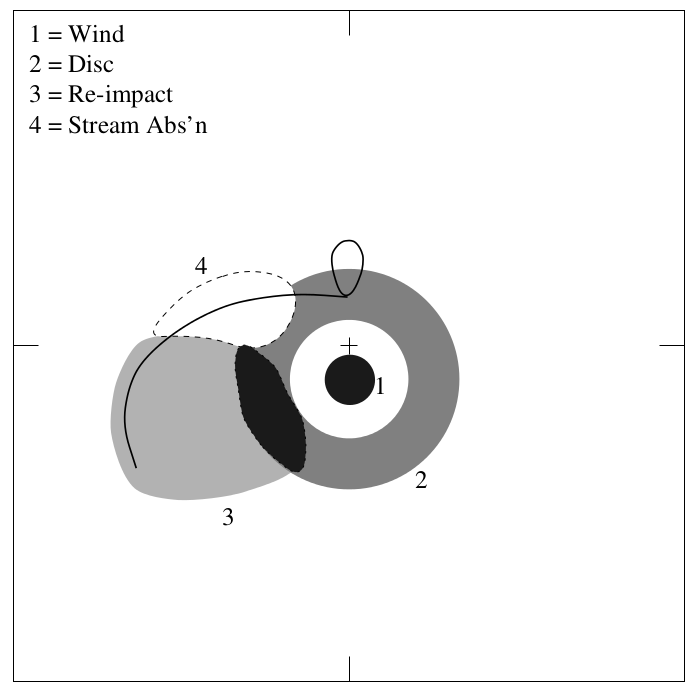}}\\[0.2cm]
\end{minipage}
}
\caption{\footnotesize The orbitally phase-binned H$\beta$ line profiles 
on the left show the absorption feature around phase 0.5. In the centre, 
the trailed spectra of H$\beta$ reveals the 
complex composition of the emission line including the high velocity
wings. For the model below, emission from a wind, the disk, and the re-impact 
site, together with P Cygni absorption and stream absorption are taken 
into account.
On the right side, the Doppler map of the H$\beta$ emission is plotted, below 
a schematic interpretation of the tomogram is given showing the position
of the various components that contribute to the emission line profile. All plots are taken from Hellier \cite{hellier96-1}. 
}
\label{fig_v1315aqldop}
\end{figure}
\begin{figure}
\centerline{
\resizebox{4cm}{!}{\includegraphics{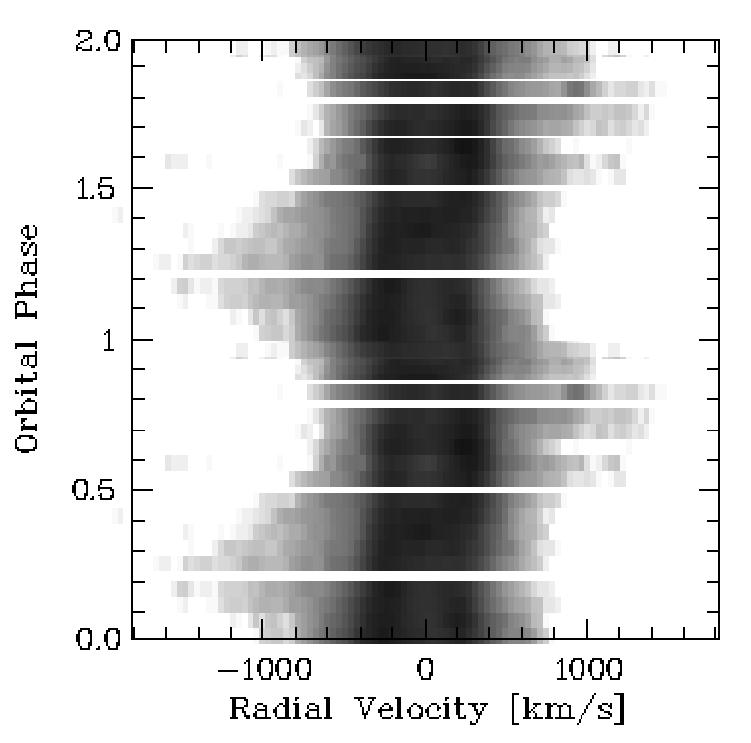}}
\resizebox{4.7cm}{!}{\includegraphics{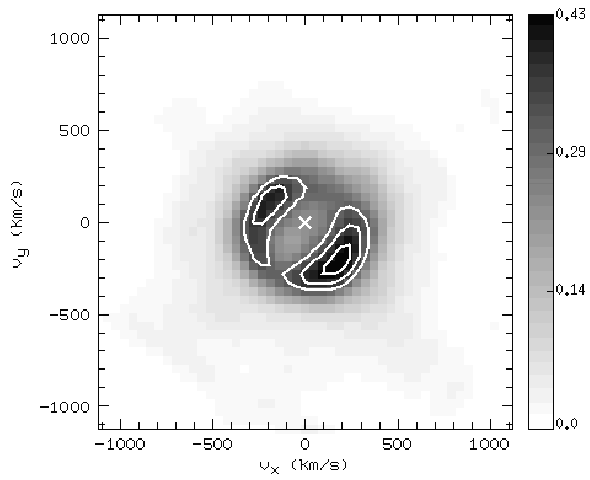}}
\resizebox{5.5cm}{!}{\includegraphics{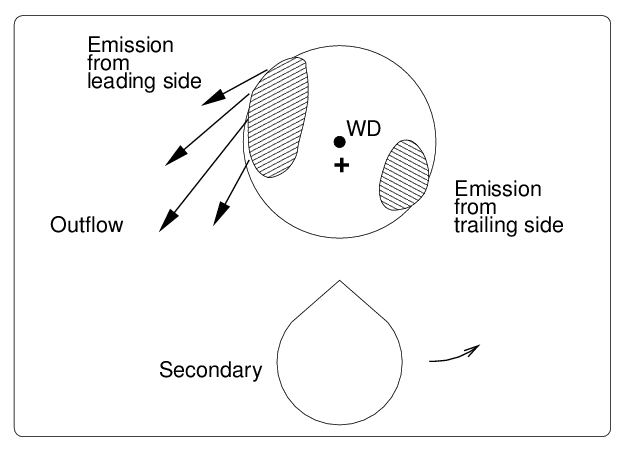}}
}
\caption{\footnotesize Trailed spectra of H$\alpha$ (left) and the 
Doppler map of the H$\alpha$ emission (centre) are plotted for RR\,Pic.
On the right side, a schematic interpretation of the data is given. 
All plots are taken from 
Schmidtobreick et al. \cite{schmidtobreicketal03-1}).
}
\label{fig_rrpic}
\end{figure}
\begin{table}
\caption{\label{tab_swsex}
The first definition of SW Sex stars as a new class of cataclysmic variables 
was done with four systems \cite{thorstensenetal91-1}. See text for details on the defining features.\smallskip
}
\begin{tabular}{|l| c| c| c| c| c| l|}
\hline
System& eclipsing & single\,peak &absorption & RV phase-lag& strong\,HeII & period/d \\
\hline
\hline
SW\,Sex & $\surd$ & $\surd$ & $\surd$ & $\surd$ & $\surd$ & 0.134938\\
DW\,UMa &  $\surd$ & $\surd$ & $\surd$ & $\surd$ & $\surd$ & 0.136607\\
V1315\,Aql &  $\surd$ & $\surd$ & $\surd$ & $\surd$ & $\surd$ & 0.139690\\
PX\,And & $\surd$ & $\surd$ & $\surd$ & $\surd$ & $\surd$ & 0.146353\\
\hline
\end{tabular}
\end{table}

After the initial classification scheme was developed,
several more systems were found
that fell into the SW\,Sex category, e.g. BH\,Lyn \cite{dhillonetal92-1},
WX\,Ari \cite{beuermannetal92-1} or V795\,Her \cite{casaresetal96-1}, 
and also two
old novae BT\,Mon \cite{smithetal98-1} 
and V533\,Her \cite{thorstensen+taylor00-1}.
At the same time, first attempts were made to understand
the SW\,Sex phenomenon and
the mechanism responsible for the mysterious features. Trailed spectra 
and Doppler maps were computed to understand the distribution of the emission
line sources, see e.g. Figure \ref{fig_v1315aqldop} and \ref{fig_rrpic}.

\begin{figure}[b]
\centerline{
\resizebox{3.8cm}{!}{\includegraphics{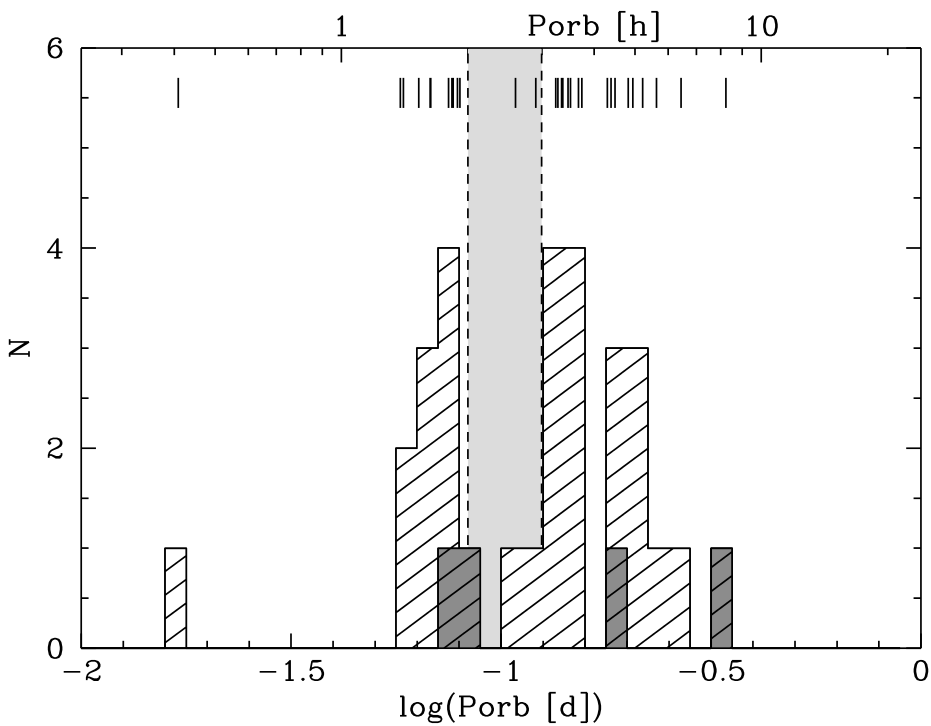}}
\resizebox{3.8cm}{!}{\includegraphics{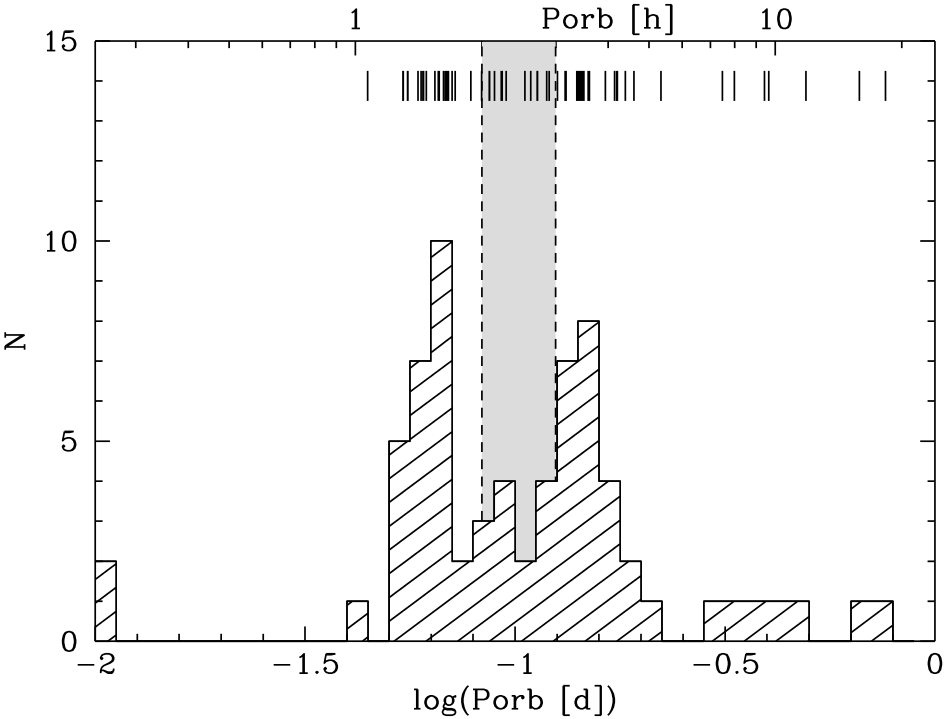}}
\resizebox{3.8cm}{!}{\includegraphics{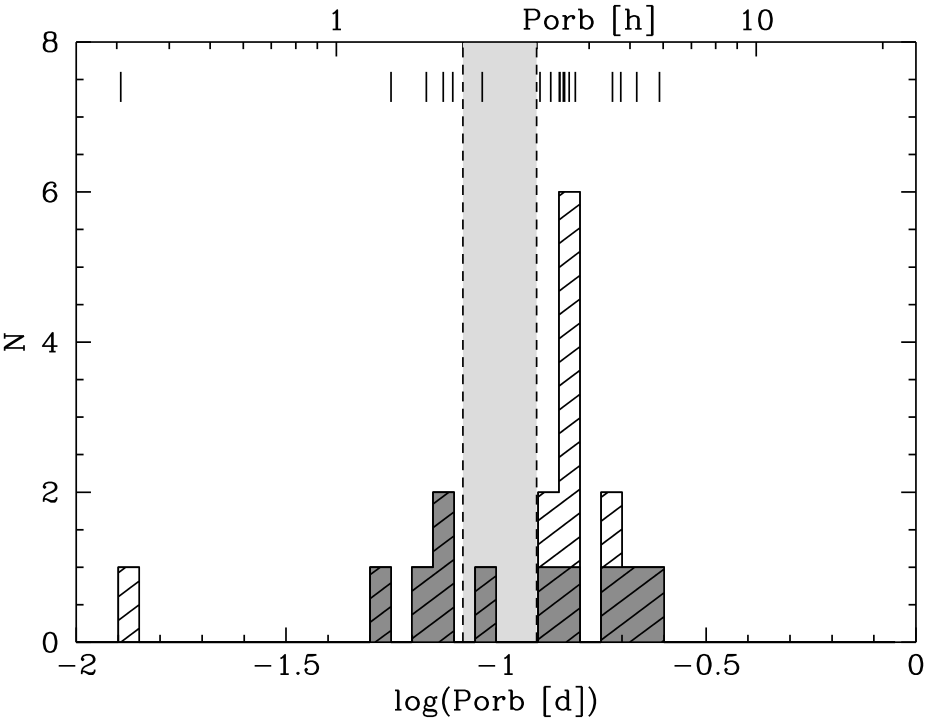}}
\resizebox{3.8cm}{!}{\includegraphics{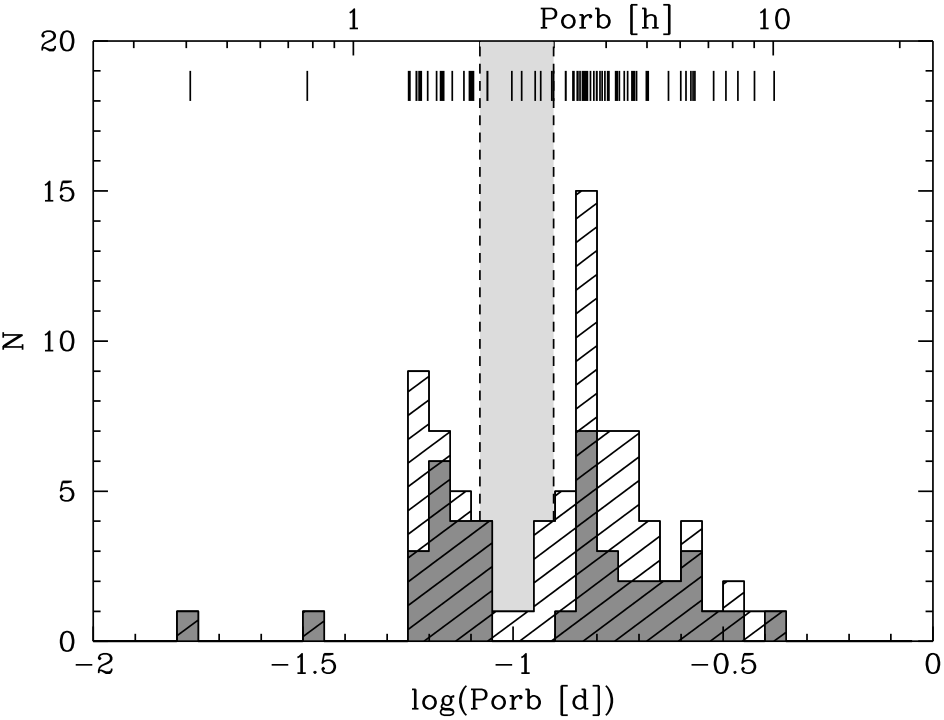}}
}
\caption{\footnotesize The orbital period distribution of CVs as found in 
various surveys from the beginning of the 21st century. From left to right:
Palomar Green, ROSAT, Edinburgh Cape and Hamburg Quasar survey.
In all plots, the period gap is indicated with a light grey bar, 
previously known 
systems are shown in dark grey. All surveys have discovered a large number of
CVs with orbital periods between 3 and 4\,h - the regime of the SW\,Sex stars 
(white bar left of the period gap).
Taken from \cite{gaensicke05-1}.
}
\label{fig_surveys}
\end{figure}

In general, SW\,Sex stars were recognised as a rare sub-species of CVs
showing peculiar behaviour. However,
already Thorstensen et al.\ 1991 found the amount
of discovered SW\,Sex systems remarkable, i.e. 3/30 in the PG survey, especially
as they have to be seen at high inclination to be discovered as such.
They noted that 'the
weird absorption events, the phase-displaced emission lines,
and so on, must be regarded as normal rather than pathological'
\cite{thorstensenetal91-1}. And indeed, with all the surveys coming 
up in the new century, the number of newly found SW\,Sex candidates 
exploded (see Figure \ref{fig_surveys} for a summary of the surveys). 
In particular, all eclipsing novalike stars with an orbital period between 
3 and 4 hours, are of SW\,Sex nature (G\"ansicke 2005 \cite{gaensicke05-1}). 
However, whether a binary is eclipsing 
or not, depends on its inclination but not on its physical properties. Thus, it
is reasonable to assume that all non- or weakly-magnetic CVs with orbital 
periods between 3 and 4 hours share the physical properties of SW\,Sex stars.
Follow-up observations of non-eclipsing CVs in this period range showed indeed
that the large majority of these systems present at least some of the defining 
SW\,Sex features  
(\cite{rodriguez-giletal07-1}, 
\cite{rodriguez-giletal07-2}, 
\cite{schmidtobreicketal12-2}). 
Rather than some exotic sub-species of CVs, SW\,Sex stars turn out 
to be the dominant population in the 3-4\,hour period range. At the 
writing of this paper, a total of 73 CVs 
 \footnote{See D.~W.\ Hoard's Big List of SW Sextantis Stars at
\tt \href{http://www.dwhoard.com/biglist}{www.dwhoard.com/biglist} \cite{hoardetal03-1}}
are considered to be possible SW\,Sex stars.

\section{The physics of the SW\,Sex features}
\begin{figure}[t]
\centerline{
\resizebox{7.5cm}{!}{\includegraphics{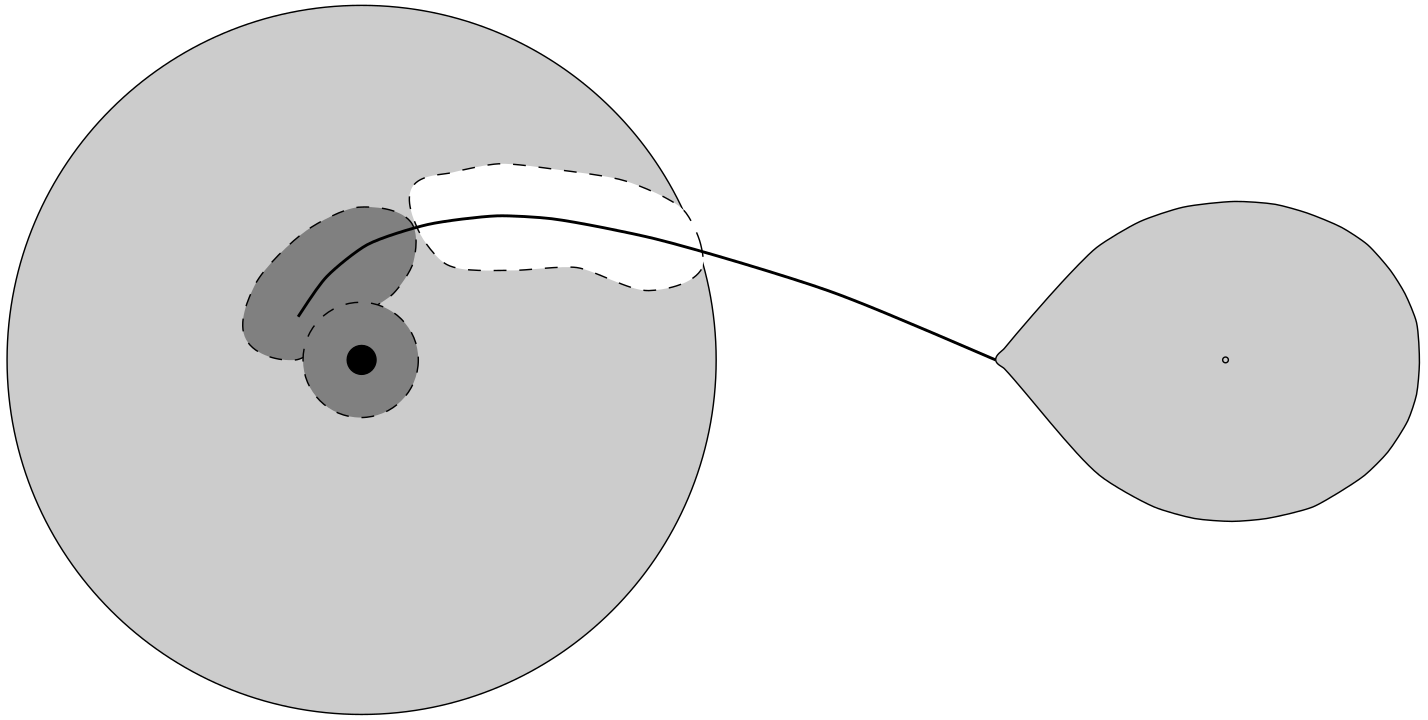}}
}
\caption{\footnotesize A schematic illustration of Hellier's model for 
SW Sex stars with a  disk overflow causing absorption (white region)
and then emission (dark region). 
An accretion disk wind from near the white dwarf adds to the emission,
filling in the double-peaked profile of the accretion disk.
\cite{hellier96-1}
}
\label{fig_hellier_mod}
\end{figure}

Already at the very beginning of the SW\,Sex studies, people were wondering
about the physical explanation for the various SW\,Sex phenomena. 
The reason, SW\,Sex stars had been identified as a distinctive class, was 
that several of the observed characteristics contradicted the general 
understanding of how CVs were built-up.
Honeycutt et al.\ \cite{honeycuttetal86-1} discussed the transient
absorption as an effect due to a somewhat discontinuous s-wave
or to the appearance of an absorption component near inferior
conjunction. To explain both, the transient absorption and the 
single-line profiles in an eclipsing binary, 
Hellier \cite{hellier96-1} developed a model of 
a disc overflow together with an accretion wind that would fill in the
double-peaked line profile from the accretion disc (see Figure \ref{fig_hellier_mod}). He suggested that such an overflow and wind could be the result of 
a high mass-transfer rate.

There are indeed more indications that SW\,Sex stars are CVs that
experience extremely high mass transfer rates.
Already Thorstensen et al.\ noted in 1991 that
'the characteristics of the SW Sex stars suggest that, except
for occasional low states, they are persistently bright
novalike variables. All were discovered in magnitude-limited
searches, which are skewed toward high intrinsic luminosity.
Thus the SW Sex phenomenon is probably very common
among the cataclysmics with the highest mass transfer
rate.' \cite{thorstensenetal91-1}
Large mass accretion rates lead to very hot white dwarfs which have been
observed e.g.\ 
by G\"ansicke et al.,\ \cite{gaensickeetal99-1}, 
Hoard et al.\ \cite{hoardetal04-1}, and 
Araujo-Betancor et al. \cite{araujo-betancoretal05-1}).
In fact, as Townsley \& G\"ansicke point out, no traditional model of 
angular momentum loss due to magnetic braking can account for 
mass accretion rates high enough to explain the observed 
white dwarf temperatures
\cite{townsley+gaensicke09-1}. 

\begin{figure}[b]
\centerline{
\resizebox{7.5cm}{!}{\includegraphics{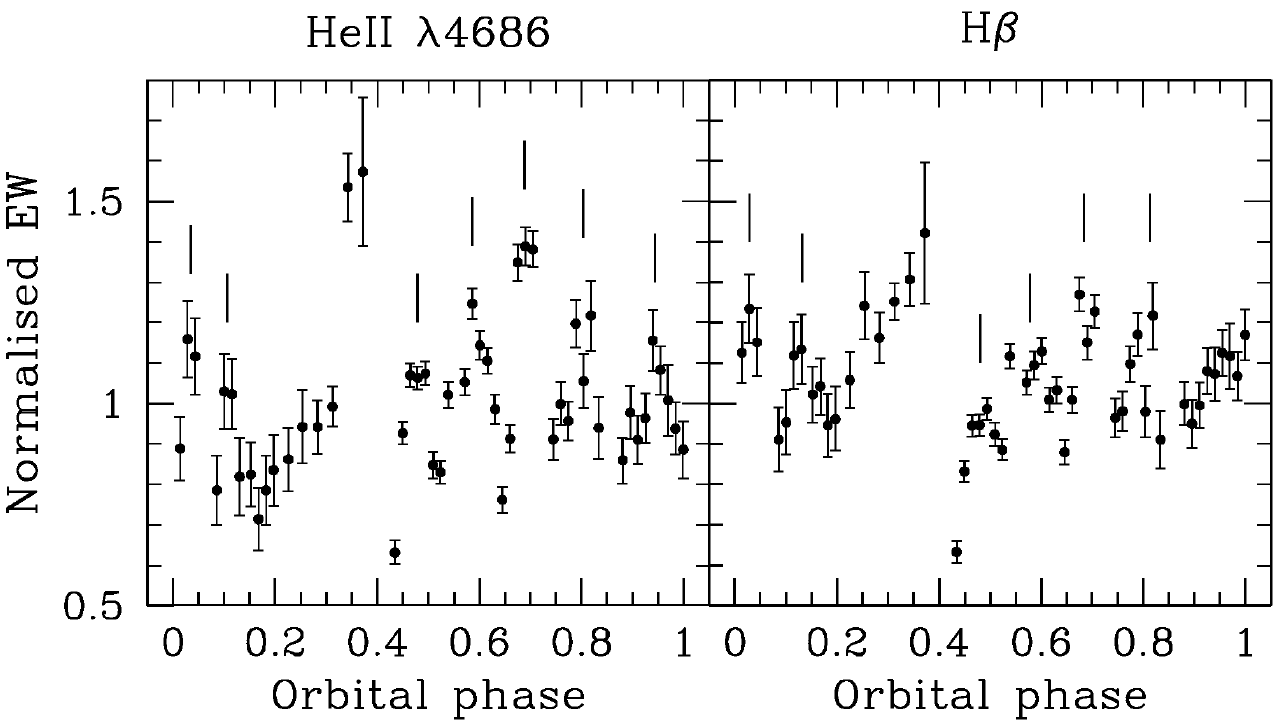}}
\resizebox{7.5cm}{!}{\includegraphics{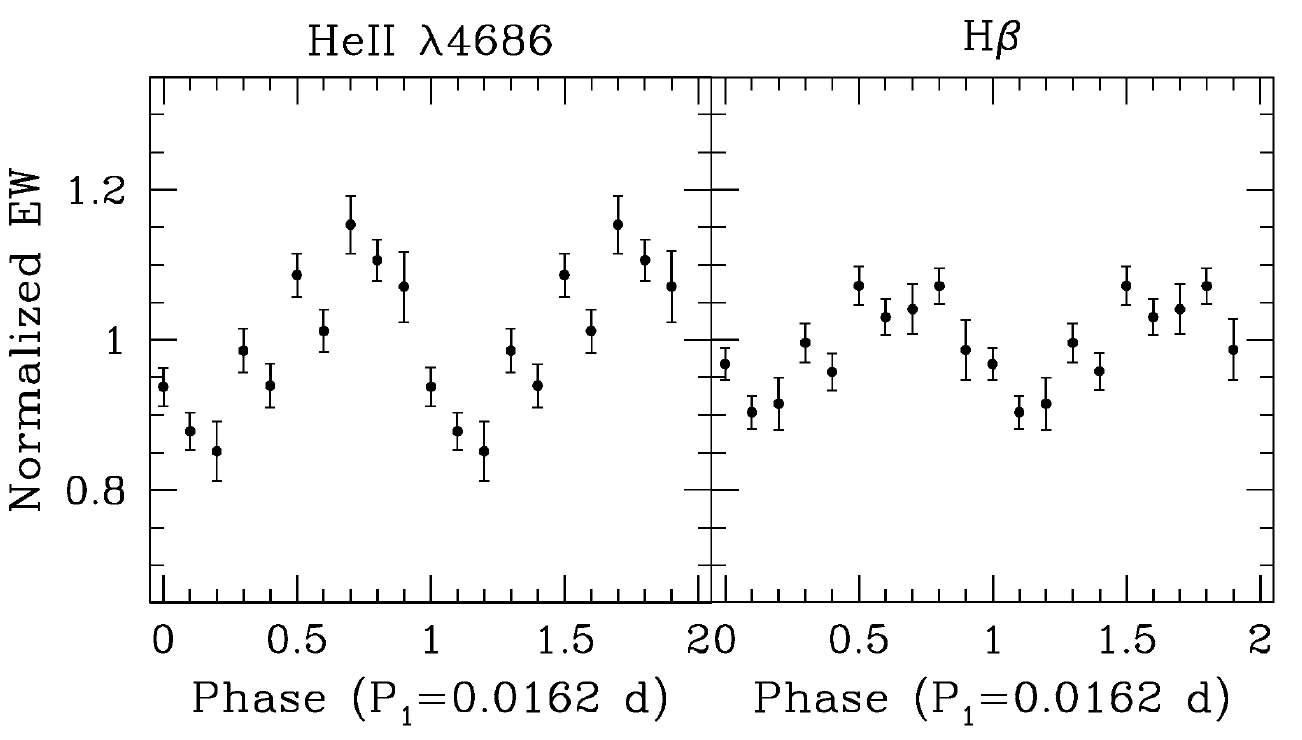}}
}
\caption{\footnotesize On the left side, the normalised equivalent widths of 
He\,II\,$\lambda$4686 and H$\beta$ are plotted against the orbital phase.
The flares are indicated with a vertical mark. On the right side,  the same
data are folded on the 23.33-min period to emphasis the periodicity of the 
flaring.
All plots taken from \cite{rodriguez-gil+martinez-pais02-1}.}
\label{fig_v533her_flares}
\end{figure}

Another puzzling feature that has been found in several
SW\,Sex stars is emission-line flaring.
Rodr\'\i guez-Gil \& Mart\'\i nez-Pais 
\cite{rodriguez-gil+martinez-pais02-1} show that the flares in the
SW\,Sex star V533\,Her are periodic
(see Figure \ref{fig_v533her_flares}). They thus interpret them
as coming from a spinning white dwarf and take it as a evidence for
the magnetic nature of the CV. More evidence for the presence of significant
magnetism in some SW\,Sex stars comes e.g.\ from the detection of variable circular
polarisation \cite{rodriguez-giletal05-1} 
and variable X-ray emission 
\cite{baskilletal05-1} in LS\,Peg.
Hoard et al.\ \cite{hoardetal03-1} discuss several physical models 
for SW\,Sex stars, including simple and complex disc overflow, but also
a stream-fed intermediate polar scenario with a truncated inner disc. In 
this picture,  SW Sex stars would be the intermediate polars with the 
highest mass transfer rates and/or the weakest magnetic fields.

\begin{figure}[t]
\centerline{
\resizebox{7.9cm}{!}{\includegraphics{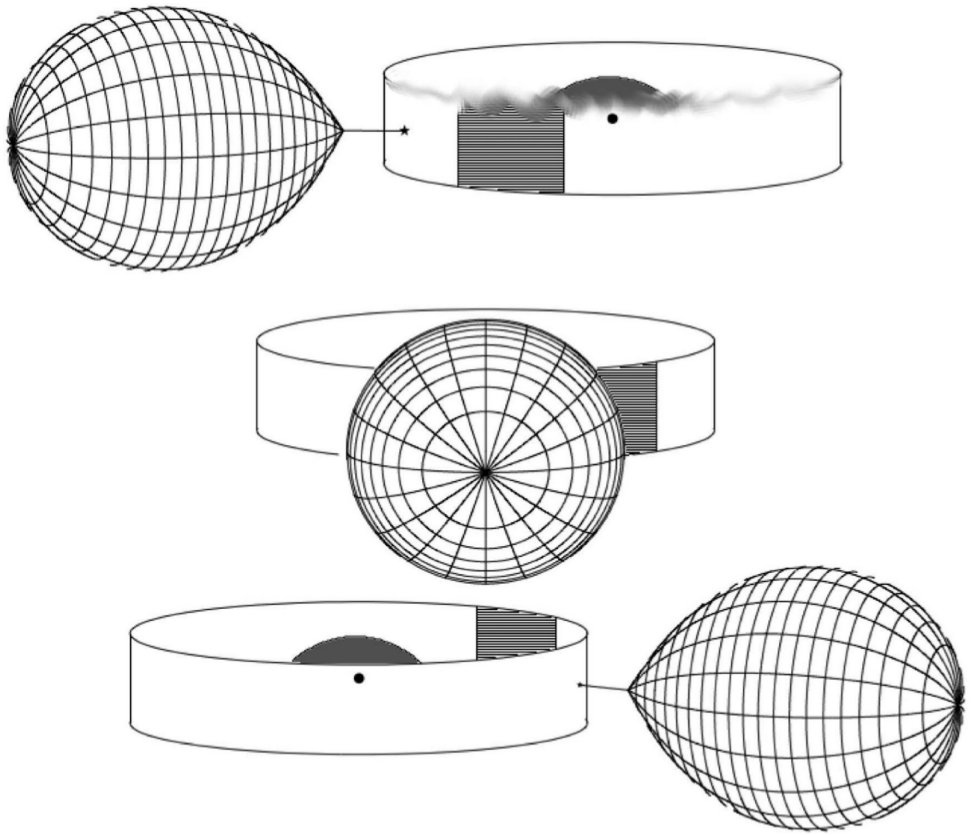}}\hspace{0.1cm}
\resizebox{6.9cm}{!}{\includegraphics{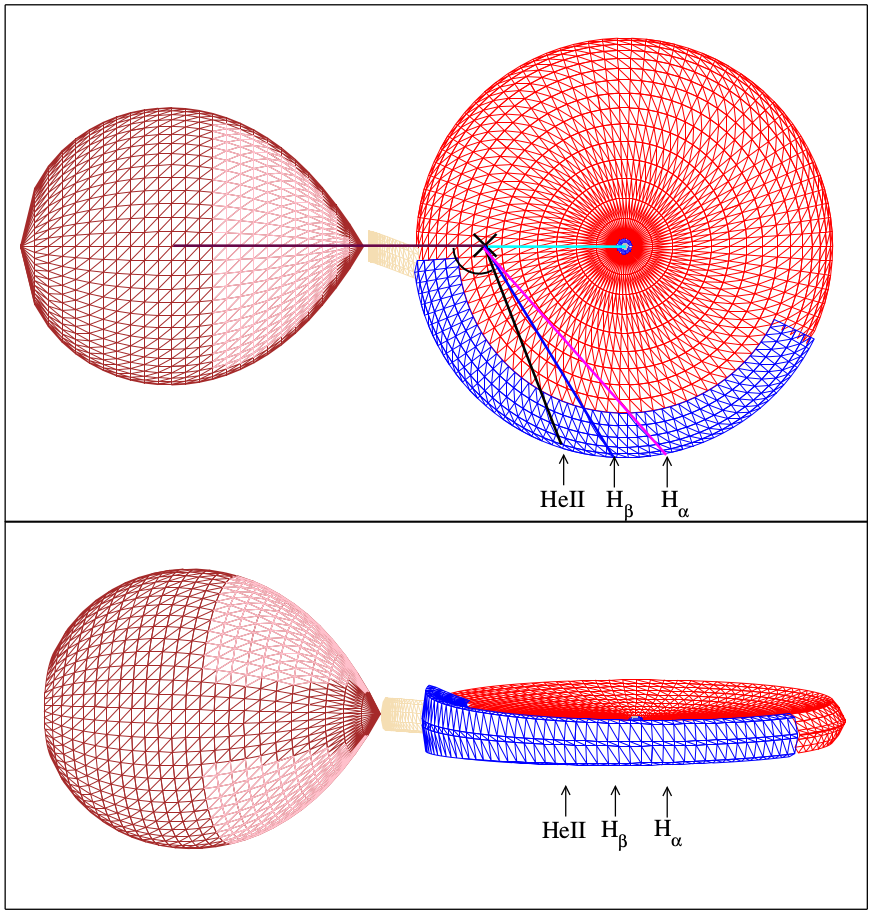}}
}
\caption{\footnotesize On the right side, a schematic model for DW\,UMa 
has been derived by Dhillon et al.\ \cite{dhillonetal13-1}. The white 
dwarf is hidden by the thick accretion disc. Key components 
of this model are the dominant Balmer emission region downstream
of the bright spot on the disc rim (shaded box), the He\,II
emission region close to the white dwarf (shaded arc) and the non-uniform
disc edge. 
Tovmassian et al.\ \cite{tovmassianetal14-1}
developed the geometric model on the right side: 
The SW Sex system consists of an 
optically thick (and thus invisible) disc and an extended hot spot 
region (blue) which is responsible for the single-peaked emission lines. 
Because of a temperature gradient in the arc, the lines of different elements
are emitted in regions that revolve around the centre of mass with different 
velocity and phase. 
}
\label{fig_tovmassian_model}
\end{figure}

More recent attempts to explain the SW\,Sex properties use simpler models and 
forgo the necessity for an out-of-plane component. Assuming an optically
thick disc as is observed also in  UX\,UMa stars, Dhillon et al.\ 
\cite{dhillonetal13-1} and Tovmassian et al.\ \cite{tovmassianetal14-1}
explain the single-peaked emission lines as they
origin not from the disc (which is optically thick and thus 
contributes to the continuum but is almost invisible
in emission lines) but from an elongated bright spot region where the
stream impacts the disc. The presence of an optically thick disc has been
confirmed by Knigge et al.\ \cite{kniggeetal04-1} from low-state
observations of DW\,UMa where the white dwarf becomes visible and changes
the slope of the spectral energy distribution in the UV. More detail
on these observations are given in section \ref{sec_low_state}.
It remains to mention that the scenario of the emission coming from
the bright spot also explains the phase delay observed in all SW\,Sex stars. 
By making the edge of the optically thick disc non-uniform, Dhillon et al. 
could also explain the emission-line flaring in DW\,UMa \cite{dhillonetal13-1}. 
Since they find no periodicity in the line flaring of this system, they rule out a 
magnetic origin of the flaring in this case.

\section{SW\,Sex stars in low state}
\label{sec_low_state}
Several SW\,Sex stars are occasionally found in states of greatly 
diminished brightness, so-called low states or VY-Scl states.
During these quiescent phases, the stars are on average
3-5 mag fainter, and they can stay at 
this level for days, months or even years before returning to the 
normal high state level. The VY-Scl behaviour is not exclusive to
SW\,Sex stars and also seems to be independent of the magnetic 
field of the white dwarf. It is observed in almost all
polars, some intermediate polars, and many weakly-magnetic CVs
like novalikes, dwarf novae, and Z Cam stars.
For reviews on such low states in CVs, see
\cite{king+cannizzo98-1}, \cite{warner99-1}, and \cite{hessman00-1}.

\begin{figure}[b]
\resizebox{0.5\hsize}{!}{
\includegraphics[clip=true]{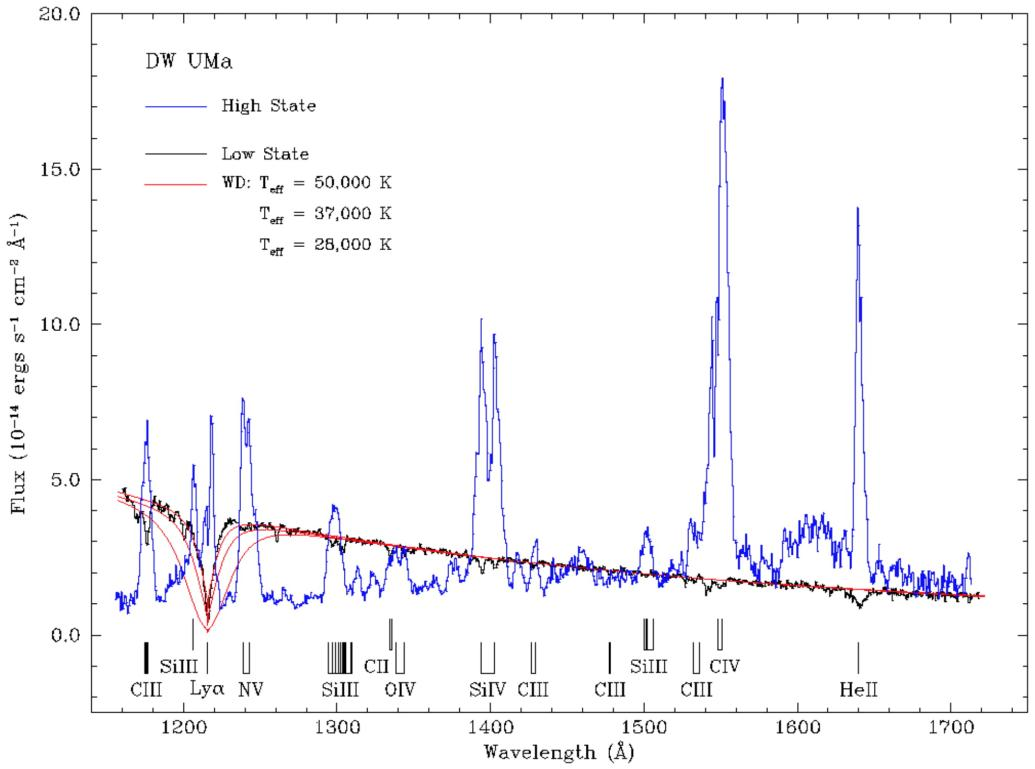}}
\resizebox{0.5\hsize}{!}{
\includegraphics[clip=true]{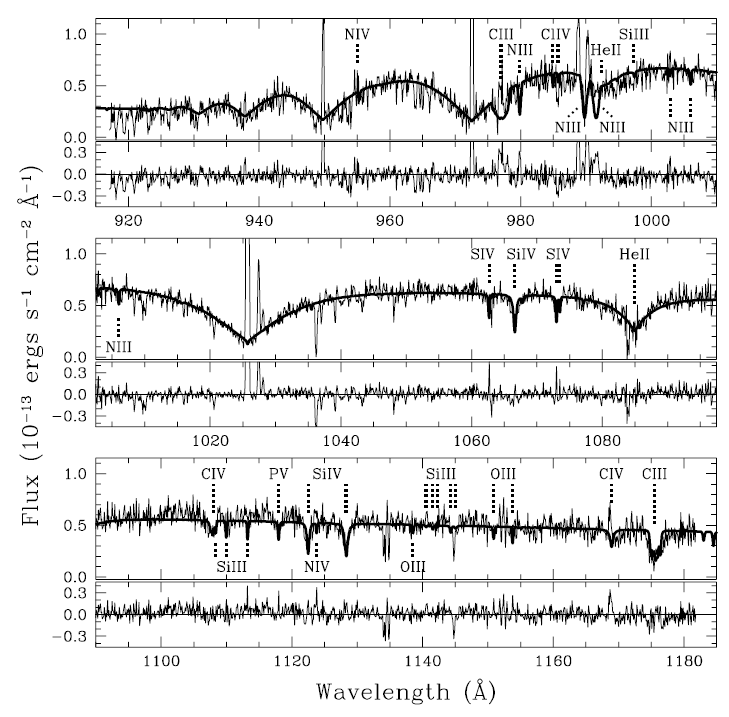}}
\caption{\footnotesize Two examples show how the features of the
white dwarf become visible during low state: On the left side,
out-of-eclipse UV spectra in the high and low state are plotted for
DW\,UMa. They demonstrate the increased blue flux in low state coming
from the white dwarf which in high state is absorbed by the accretion disc \cite{kniggeetal04-1})
On the right side, the far UV spectrum of MV\,Lyr is plotted; broad absoption loines from the white dwarf and some narrow emission lines are visible \cite{hoardetal04-1}. 
}
\label{fig_lowstate}
\end{figure}

It is widely accepted, that during such low states, the mass transfer 
from the donor star through the L1 point is reduced or even completely
suppressed, even though the exact mechanism for this cessation of mass 
supply from the secondary remains unclear. 
Livio \& Pringle \cite{livio+pringle94-1}
discuss that Roche-lobe overflow can be inhibited by the accumulation of 
starspots on L1 and this idea is also supported by 
King \& Cannizzo \cite{king+cannizzo98-1}. 
Using Roche-Tomography techniques, Watson et al.\
\cite{watsonetal06-1} \cite{watsonetal07-1} demonstrated the 
presence of starspots on the donor stars of CVs and showed that they 
cluster on the side facing the white dwarf. 
Hoard et al. \cite{hoardetal04-1}) observed MV\,Lyr during 
low state and found evidence for the possible presence of starspots 
on the secondary star near the L1 point. A different approach to explain
the phases of subdued mass transfer is taken by Wu et al.\ 
\cite{wuetal95-3} who discuss the unstable mass transfer to be
caused by an interplay of irradiation, heating and disc shielding of 
the donor star.

Whatever the reason, mass transfer ceases and the accretion disc 
decreases in size or disappears completely and the overall system brightness
decreases.
The main advantage of observing CVs in 
low states is this weakness or absence of the accretion disc which, 
especially for SW\,Sex stars, is normally the dominating light source in 
the binary and which is then
veiling the emission from the stellar components. Thus, low states provide 
a unique opportunity to study the white dwarf or the donor star in 
these systems, to determine system parameters such as temperature, 
mass, radius, and stellar types from the spectrum 
and - by following their radial velocity curves over time - 
to derive the dynamical masses of these components. 

Araujo-Betancor et al. \cite{araujo-betancoretal03-1} analysed  
ultraviolet (UV) eclipse data of DW\,UMa that were obtained during a 
low state in which the UV light was dominated by the hot white dwarf primary.
From these, they put new constraints on the binary parameters and the
stellar components.
Knigge et al. \cite{kniggeetal04-1}
confirm that the UV flux in the low state is higher and bluer than in the
high state. This is explained by assuming that in high state, an optically 
thick accretion disc absorbs the UV emission of the white dwarf (see Figure \ref{fig_lowstate}, left side).
\begin{figure}[b]
\resizebox{0.43\hsize}{!}{
\includegraphics[clip=true]{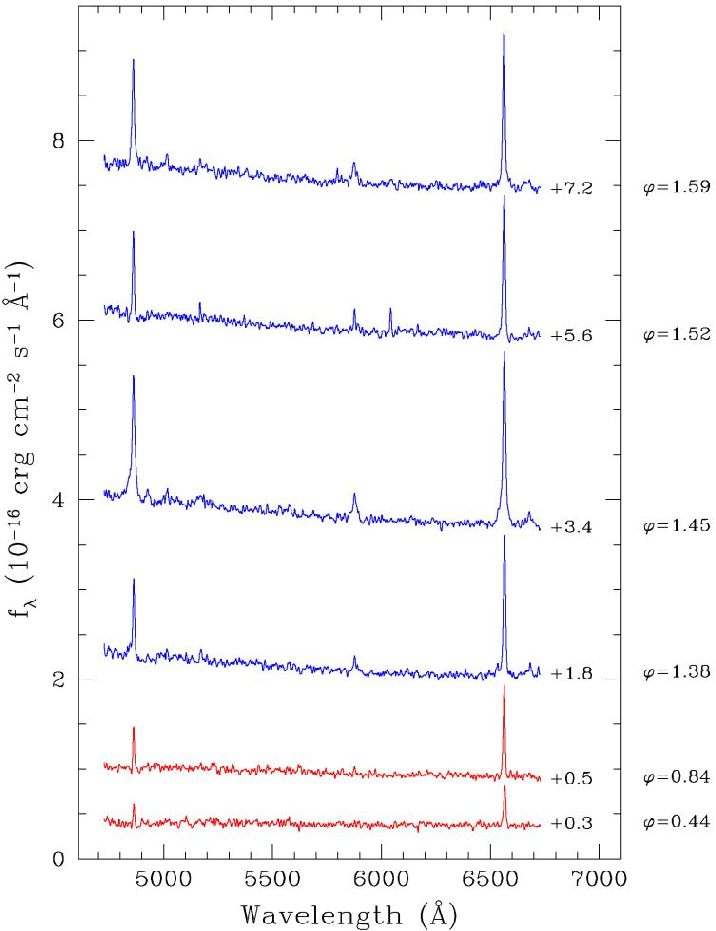}}
\hspace{0.2cm}
\resizebox{0.55\hsize}{!}{
\includegraphics[clip=true]{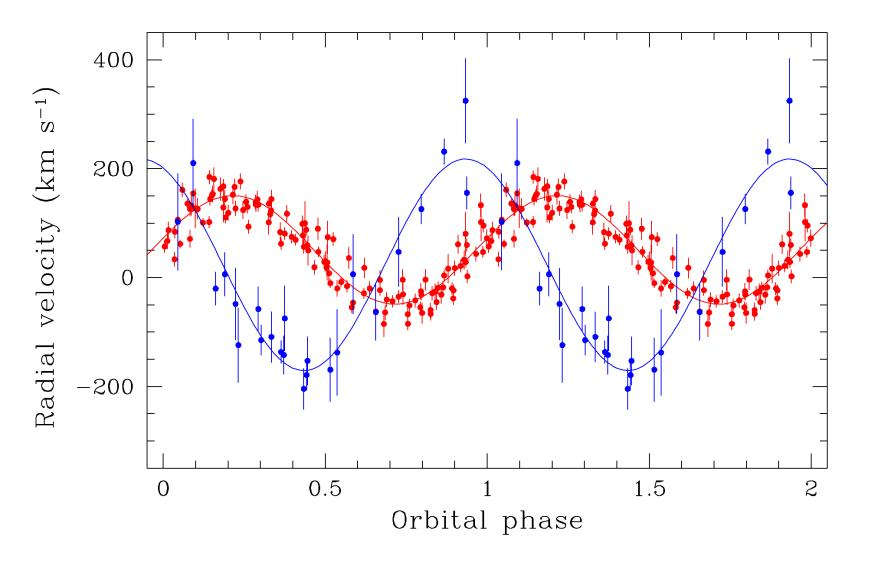}}
\caption{\footnotesize
The sporadic mass transfer events in the low state of BB\,Dor are demonstrated:
On the left, an observing sequence of about
four hours of is plotted, on the right the
H$\alpha$ radial velocities in quiescence (red dots) and during the
accretion events (blue dots). The solid lines are the
respective best sine fits. The phases of the radial velocities from the
quiescent spectra indicate that the origin lies in the irradiated
donor star.  The radial velocities during the accretion events are delayed
by 0.18 cycle with respect to the expected motion of the white dwarf and thus do not
origin in a disc around the white dwarf but rather in a stream or impact region.
\cite{rodriguez-giletal12-1}.
\label{fig_bbdorblob}}
\end{figure}

Hoard et al.\ \cite{hoardetal04-1}) obtained a far-ultraviolet spectrum
(see Figure \ref{fig_lowstate}, right side), time-resolved optical photometry, 
and optical spectroscopy
of MV\,Lyr in low state. By combining these data, they establish 
a model for the white dwarf, the secondary star and the binary system. 
They find no evidence for the presence of an accretion disc. 

\begin{figure}[t]
\resizebox{0.45\hsize}{!}{
\includegraphics[clip=true]{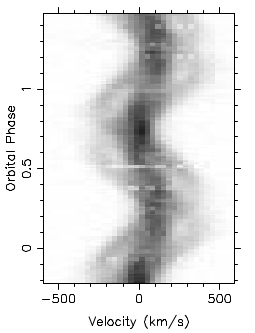}}
\hspace{0.2cm}
\resizebox{0.55\hsize}{!}{
\includegraphics[clip=true]{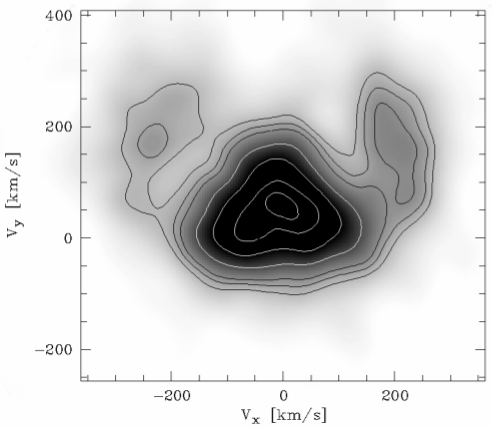}}\\[0.5cm]
\resizebox{0.4\hsize}{!}{
\includegraphics[clip=true]{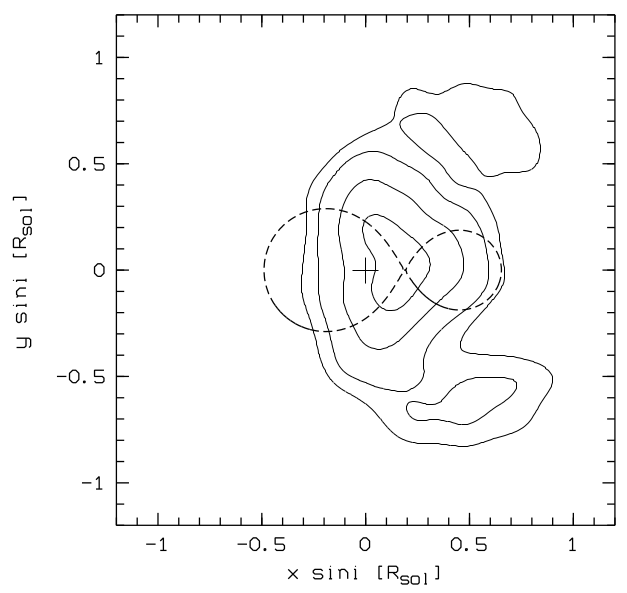}}
\resizebox{0.6\hsize}{!}{
\includegraphics[clip=true]{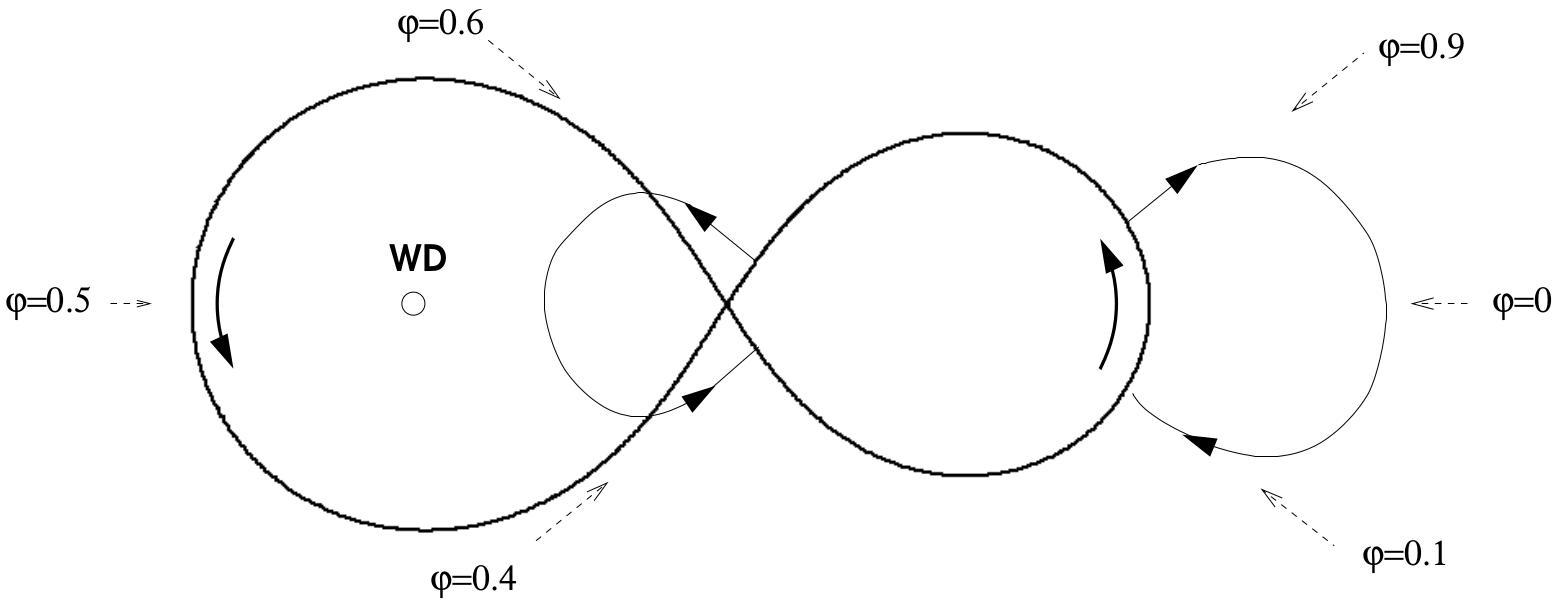}}
\caption{\footnotesize 
The four plots are taken from 
Schmidtobreick et al. \cite{schmidtobreicketal12-1} and 
show the presence and interpretation of 
satellite lines in BB\,Dor in low state. On the upper left the trailed 
spectrograms of the H$\alpha$ line are plotted. The strong line moving in 
the centre is attributed to the irradiated secondary star. The satellite lines
following two crossing sinusoidal curves are clearly visible. The upper right
plot shows the Doppler map of the  H$\alpha$ emission sources in the system.
Assuming that all emission sources are bound to the orbit, the velocities of
the H$\alpha$ emission can be converted into Cartesian coordinates 
(see lower left plot). The sketch on the lower right side
illustrates the two possible orientations of prominences that could explain 
the observed satellite lines. Scaling and sizes are arbitrary.
\label{fig_bbdorsat}}
\end{figure}

In 2008, BB Doradus faded from V$\approx 14.3$ towards a deep low state
at V$\approx 19.3$. During this low state, it was monitored photometrically 
and spectroscopically by 
Rodr\'\i guez-Gil et al.\ \cite{rodriguez-giletal12-1}
and Schmidtobreick et al.\ \cite{schmidtobreicketal12-1}.
They observed episodic accretion events which veiled the absorption spectra 
of the white dwarf and the donor and radically changed the line profiles on
timescale of tens of minutes: the narrow emission lines which originate
in the irradiated secondary switched to broader
lines with strong wings, and high-excitation lines such as He\,II and 
the Bowen blend appeared. A shift of about 0.35 phases between the 
radial velocities of the narrow lines and those of the broad line wings 
is found. This shows that accretion is not complete quenched in the low 
state but can be switched on fast, possibly triggered by flaring activity on the
secondary star \cite{rodriguez-giletal12-1}. 
Schmidtobreick et al.\ \cite{schmidtobreicketal12-1} analysed 
quiescence data taken
outside such accretion events. The spectra show narrow emission lines of 
H$\alpha$, He\,I and Na-D, as well as TiO absorption troughs
which trace the motion of different temperature zones on the 
irradiated secondary star. While  He\,I and H$\alpha$ originate at the 
higher temperature regions of the secondary star close to L1 
(which is reflected in their low radial-velocity amplitude), the TiO bands 
are likely to originate in the non-irradiated, cooler, backside zones of the
secondary star (high radial-velocity amplitude). No signature of accretion
is present in the spectra. However, they detect two additional
components in the H$\alpha$ emission line, satellite lines similar to what 
has been detected
previously in AM\,Her stars 
\cite{kafkaetal05-1} \cite{kafkaetal06-1} \cite{kafkaetal07-1} \cite{kafkaetal08-1} \cite{kafkaetal10-1} \cite{masonetal08-1}. 
A modified Roche geometry taking 
into account an additional magnetic force might result in stable equilibrium 
points around the positions where the increased emission is observed.
A different possibility is that 
the H$\alpha$ emitting material moves along prominences originating 
in a magnetically active secondary star.
To explain the observed variation of the satellite lines, only
two orientations are possible for these prominences 
(see Figure \ref{fig_bbdorsat}
for the data and a schematic interpretation)

\begin{figure}[t]
\centerline{
\resizebox{8cm}{!}{\includegraphics{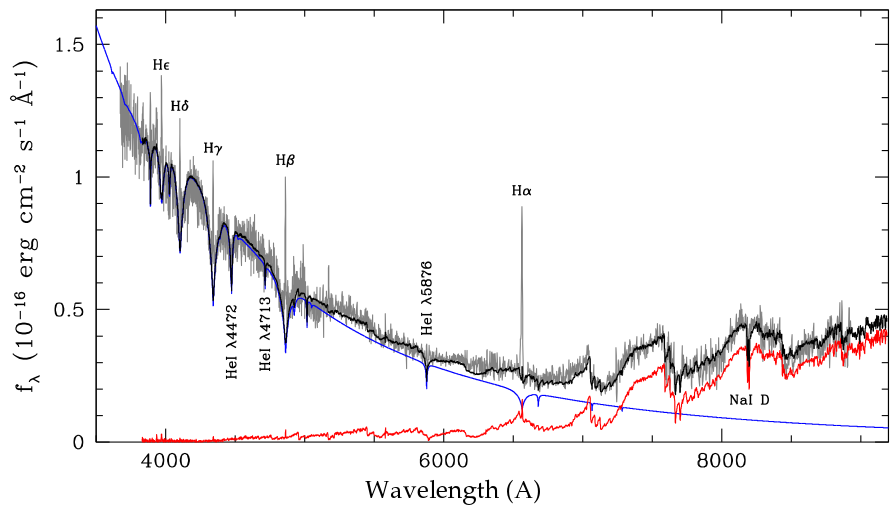}}\hspace{0.2cm}
\resizebox{6.8cm}{!}{\includegraphics{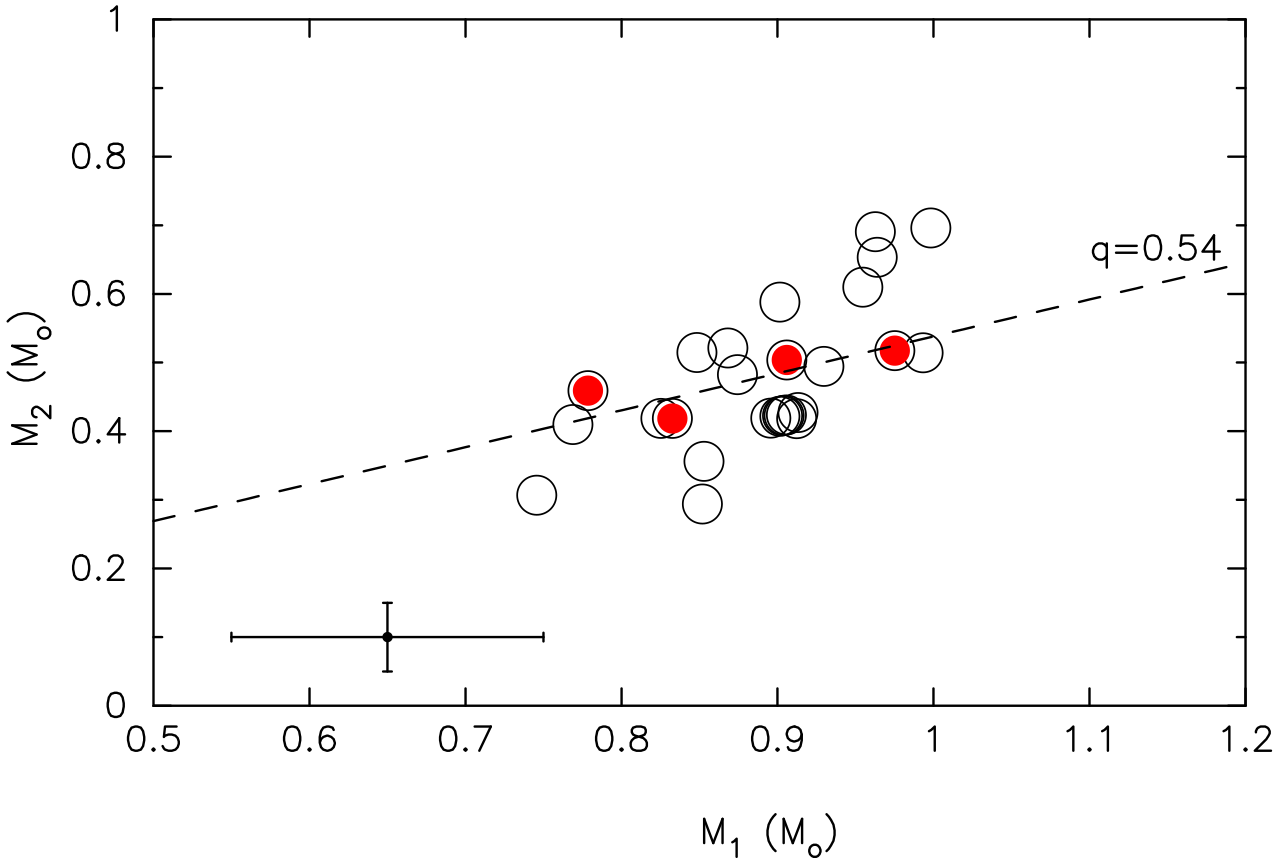}}}
\caption{\label{fig_hs0220}The left presents the average quiescence spectrum of
 HS\,0220+0603. The data (grey) are fitted with white dwarf (blue) + M-dwarf 
(red) templates.  On the right, the $M_1-M_2$ plane is plotted
with the best-fitting solutions (99 per cent confidence level for red circles)
The dashed line corresponds to a mass ratio of q = 0.54. \cite{rodriguez-giletal15-1}}
\end{figure}

The last example of an analysis of low-state data of a SW\,Sex system
refers to the eclipsing star HS\,0220+0603 which was extensively observed
during the  2004-2005 low state by Rodr\'\i guez-Gil et al. (2015)
\cite{rodriguez-giletal15-1}. Their dataset include time-resolved spectroscopy 
and photometry in various optical bands. Both the white dwarf and the secondary
star are visible in the optical spectra. Spectral modelling revealed 
the secondary to be of type M5.5 with an effective temperature of 2800\,K. 
The white dwarf turned out to be of DAB type and with an effective 
temperature of 30.000\,K. Simultaneously modelling of the eclipse 
lightcurve and the radial velocities of both components, allowed to
derive dynamical stellar masses for the first time (Figure \ref{fig_hs0220}). 

\section{SW\,Sex stars as an evolutionary state}
In general, the evolution of stable contact binaries is driven by the 
angular momentum loss which as such determines the mass transfer rate.
This implies that the general evolutionary direction for CVs is from 
longer to shorter orbital periods. 
According to the standard model of CV evolution, the main source of 
angular momentum loss for CVs with orbital periods $P_\mathrm{orb} > 3$\,h 
is the magnetic braking. Due to the continuous mass transfer, the
secondary is pushed out of thermal equilibrium and becomes bloated. 
At a period of about 3 h, i.e. at the upper edge of the period gap, the 
secondary becomes fully convective, magnetic braking ceases and
only the much weaker braking via gravitational radiation 
continues. This allows the star to approach thermal equilibrium and to shrink 
to the volume corresponding to its mass. It thus loses contact with its Roche lobe, mass transfer stops and the CV enters the period gap.
For an extensive review of the current 
understanding of CV evolution, see Knigge et al.\ \cite{kniggeetal11-1}.

Given their position in the period distribution, CVs with
orbital periods between 3\,h and 4\,h thus play a major role in our 
understanding of CV evolution. They are the ones at the 
upper edge of the period gap whose donor stars  
are predicted to become fully convective and where magnetic braking
is about to cease so they  will become a detached CV
in the next future. Rodr\'{\i}guez-Gil et al. 
\cite{rodriguez-giletal07-1} 
have shown that at least 50\% of all CVs in this period regime are
SW\,Sex stars, novalikes with extremely
high mass accretion rates leading to very hot white dwarfs.
As mentioned
earlier, no angular momentum loss by traditional magnetic braking can 
account for such high mass transfer rates and white-dwarf temperatures 
\cite{townsley+gaensicke09-1}. So the question
arises whether this unusual high mass transfer rate that manifests
itself just for these binaries where major changes in the donor structure
are supposed to happen, is in any way connected to these changes. 
As far as I am aware, only one approach has been made to explain the
high mass transfer rate above the period gap in terms of CV evolution.
Zangrilli et al.\ \cite{zangrillietal97-1} assume that for the
CVs above the period gap, only the convective envelope of the donor is 
co-rotating with the binary while the radiative core remains at a lower
rotation velocity. They calculate the angular momentum loss arising from
 a combination of two $\alpha-\Omega$ dynamos, one 
in the envelope and one at the boundary between core and envelope. By the
time the radiative core disappears, the boundary dynamo becomes the
dominant source of magnetic braking and would the mass transfer rates
for those CVs that inhabit the upper edge of the period gap, the SW\,Sex stars.

\begin{figure}[t]
\centerline{
\resizebox{12cm}{!}{\includegraphics{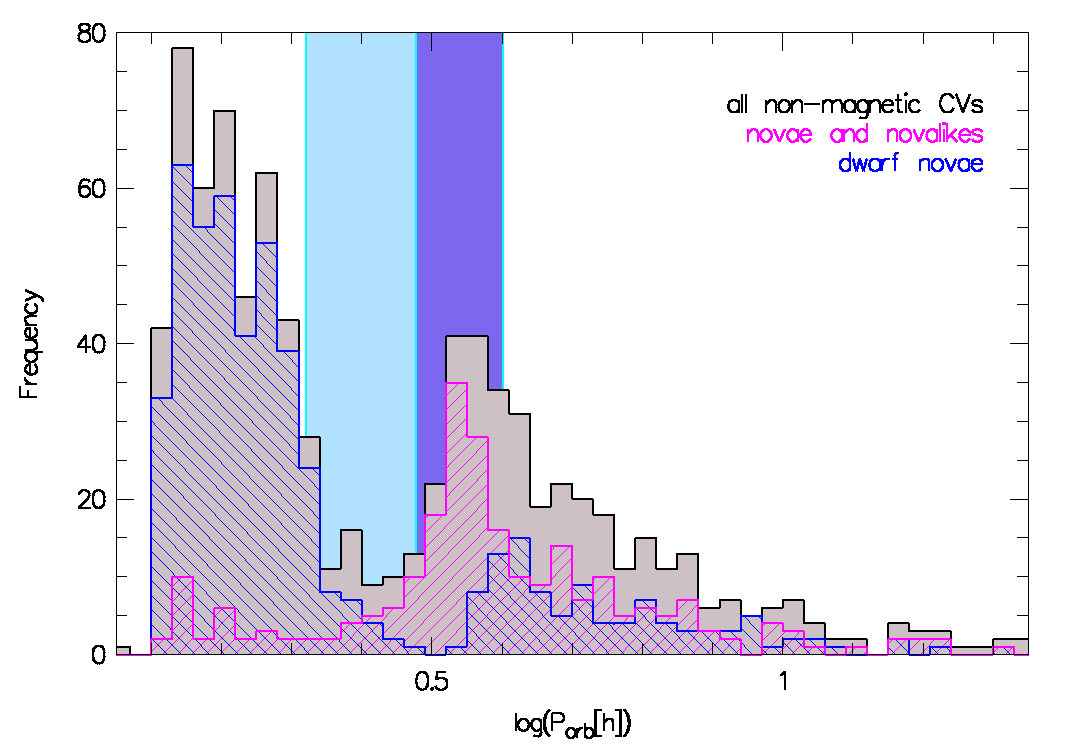}}}
\caption{\label{fig_periods} The orbital period distribution for 
different subtypes of non-magnetic CVs has been
built using the data from Ritter \& Kolb (2003), update RKcat7.18, 2012.
The position of the period gap is indicated by the light blue bar, the darker
bar indicates the 3-4\,h range dominated by systems with high mass transfer.
The number of dwarf novae decreases at the upper edge of the SW-Sex regime.
}
\end{figure}

Schmidtobreick et al.\ \cite{schmidtobreicketal12-2} demonstrate that
all the novalike-stars in their sample of non-magnetic CVs with periods
between 3 and 4 hours that have sufficient data observed, show
at least some SW\,Sex characteristics. They argue that the 
nova-like population in the 3-4\,h period regime is clearly dominated by 
SW Sex stars and almost all CVs in this period regime show the extremely 
high mass transfer rates. Since according 
to the standard model, all long period CVs have to evolve
through this period regime before entering the period gap, they should then
 share the SW\,Sex characteristics during that time.
In fact, the 
orbital period distribution shows a decrease of dwarf novae around 4\,h
just where the SW\,Sex stars begin to appear (see Figure \ref{fig_periods}). 
One can assume that their 
mass transfer ceases earlier than for nova-like stars or that they evolve 
into the high mass-transfer systems. In either case, the dominant SW\,Sex 
phenomenon between 3\,h and 4\,h is a phase in the secular evolution of 
CVs that challenges the standard model of CV evolution. 
\section{Summary}
When they were originally discovered, 
SW\,Sex stars were considered peculiar systems,
characterised by enigmatic properties that
disagreed with the general understanding of CVs and their components. 
During the last decades it
became, however, clear that these systems represent a large fraction of CVs 
and actually dominate the population at the upper edge of the period gap.
The SW\,Sex characteristics can be explained by 
assuming high mass transfer rates leading to hot accretion
streams, optically thick discs, and outflows that in combination produce the
observed features. The definition of SW\,Sex stars has become broader, aiming
more for the physical difference rather then the phenomenology. Thus, we now 
consider the SW\,Sex stars as those nova-like CVs with the most extreme mass 
transfer rates. 
The reason why they accumulate in the
orbital period range above the gap is still unknown, but a connection
to the secular evolution of CVs in this range seems likely.


\begin{thebibliography}{99}
\bibitem{araujo-betancoretal05-1}
{Araujo-Betancor} S., {G{\" a}nsicke} B.~T., {Hagen} H.-J., {Marsh} T.~R., {Harlaftis} E.~T., {Thorstensen} J., {Fried} R.~E., {Schmeer} P., {Engels} D., 2005, A\&A 430, 629
\bibitem{araujo-betancoretal03-1}
{Araujo-Betancor} S., {Knigge} C., {Long} K.~S., {Hoard} D.~W., {Szkody} P., {Rodgers} B., {Krisciunas} K., {Dhillon} V.~S., {Hynes} R.~I., {Patterson} J., {Kemp} J. 2003, ApJ 583, 437
\bibitem{baskilletal05-1}
{Baskill} D.~S., {Wheatley} P.~J., {Osborne} J.~P., 2005, MNRAS 357, 626
\bibitem{beuermannetal92-1}
{Beuermann} K., {Thorstensen} J.~R., {Schwope} A.~D., {Ringwald} F.~A., {Sahin} H. 1992, A\&A 256, 442
\bibitem{casaresetal96-1}
{Casares} J., {Mart\'\i nez-Pais} I.~G., {Marsh} T.~R., {Charles} P.~A., {L\'azaro} C. 1996, MNRAS 278, 219
\bibitem{dhillon90-1}
{Dhillon} V.~S., 1990, PhD Thesis
\bibitem{dhillonetal92-1}
{{Dhillon} V.~S., {Jones} D.~H.~P., {Marsh} T.~R., {Smith} R.~C.} 1992, MNRAS 258, 225
\bibitem{dhillonetal13-1}
{{Dhillon} V.~S., {Smith} D.~A., {Marsh} T.~R. 2013, MNRAS 428, 3559
\bibitem{gaensickeetal99-1}
{G\"ansicke} B.~T., {Sion} E.~M., {Beuermann} K., {Fabian} D., {Cheng} F.~H., {Krautter} J., 1999, A\&A 347, 178
\bibitem{gaensicke05-1}
{G\"ansicke}  B.~T. 2005, ASP Conf.~Ser.~330, 3
\bibitem{hellier96-1}
{Hellier} C. 1996, ApJ 471, 949
\bibitem{hessman00-1}
{Hessman} F.~V., 2000, NewAR 44, 155
\bibitem{hoardetal04-1}
{Hoard} D.~W., {Linnell} A.~P., {Szkody} P., {Fried} R.~E., {Sion} E.~M., {Hubeny} I., {Wolfe} M.~A. 2004, ApJ 604, 346
\bibitem{hoardetal03-1}
{Hoard} D.~W., {Szkody} P.,  {Froning} C.~S., {Long} K.~S.{Knigge} C. 2003, AJ 126, 2473
\bibitem{honeycuttetal86-1}
{Honeycutt} R.~K., {Schlegel} E.~M., {Kaitchuck} R.~H., 1986, ApJ 302, 388
\bibitem{horne+marsh86-1}
{Horne} K., {Marsh} T.~R.} 1986, MNRAS 218, 761
\bibitem{kafkaetal05-1}
{Kafka} S., {Honeycutt} R.~K., {Howell} S.~B., {Harrison} T.~E, 2005, AJ 130, 2852
\bibitem{kafkaetal06-1}
{Kafka} S., {Honeycutt} R.~K., {Howell} S.~B., 2006, AJ 131, 2673
\bibitem{kafkaetal07-1}
{Kafka} S., {Howell} S.~B., {Honeycutt} R.~K., {Robertson} J.~W., 2007, AJ 133, 1645
\bibitem{kafkaetal08-1}
{Kafka} S., {Ribeiro} T., Baptista, R., {Honeycutt} R.~K., {Robertson} J.~W., 2008, ApJ 688, 1302 
\bibitem{kafkaetal10-1}
{Kafka} S., Tappert C., {Ribeiro} T., {Honeycutt} R.~K., {Hoard}  D.~W., {Saar}, S., 2010, ApJ 721, 1714
\bibitem{king+cannizzo98-1}
{King} A.\ R., {Cannizzo} J.\ K., 1998, ApJ 499, 348
\bibitem{kniggeetal04-1}
{Knigge} C., {Araujo-Betancor} S., {G{\"a}nsicke} B.~T., {Long} K.~S., {Szkody} P., {Hoard} D.~W., {Hynes} R.~I., {Dhillon} V.~S. 2004, ApJ 615, L129
\bibitem{kniggeetal11-1}
{Knigge} C., Baraffe I., Patterson J., 2011, ApJS 194, 28
\bibitem{livio+pringle94-1}
{Livio} M., {Pringle} J.\ E., 1994, ApJ 427, 956
\bibitem{masonetal08-1}
Mason E., {Howell} S.~B., Barman T., Szkody P., Wickramasinghe D., 2008, A\&A 490, 279
\bibitem{ritter+kolb03-1}
{Ritter} H., {Kolb} U., 2003, A\&A 404, 301
\bibitem{rodriguez-gil+martinez-pais02-1}
{Rodr{\'{\i}}guez-Gil} P., {Mart{\'{\i}}nez-Pais} I.~G. 2002, MNRAS 337, 209
\bibitem{rodriguez-giletal05-1}
{Rodr{\'{\i}}guez-Gil} P., {Casares} J., {Mart{\'{\i}}nez-Pais} I.~G., Hakala P., Steeghs D., 2001, ApJ 548, L49
\bibitem{rodriguez-giletal15-1}
{Rodr{\'{\i}}guez-Gil} P., {Shahbaz} T., {Marsh} T.~R., {G{\"a}nsicke} B.~T., {Steeghs} D., {Long} K.~S., {Mart{\'{\i}}nez-Pais} I.~G., {Armas Padilla} M., {Schwarz} R., {Schreiber} M.~R., {Torres} M.~A.~P., {Koester} D., {Dhillon} V.~S., {Castellano} J., {Rodr{\'{\i}}guez} D. 2015, MNRAS 452, 146
\bibitem{rodriguez-giletal12-1}
{Rodr{\'{\i}}guez-Gil} P., {Schmidtobreick} L., {Long} K.~S., {G{\"a}nsicke} B.~T., {Torres} M.~A.~P., {Rubio-D{\'{\i}}ez} M.~M., {Santander-Garc{\'{\i}}a} M. 2012, MNRAS 422, 2332
\bibitem{rodriguez-giletal07-1}
{Rodr{\'{\i}}guez-Gil} P., {G{\"a}nsicke} B.~T., {Hagen} H.-J., {Araujo-Betancor} S., {Aungwerojwit} A., {Allende Prieto} C., {Boyd} D., {Casares} J., {Engels} D., {Giannakis} O., {Harlaftis} E.~T., {Kube} J., {Lehto} H., {Mart{\'{\i}}nez-Pais} I.~G., {Schwarz} R., {Skidmore} W., {Staude} A., {Torres} M.~A.~P. 2007(A), MNRAS 377, 1747
\bibitem{rodriguez-giletal07-2}
{Rodr{\'{\i}}guez-Gil} P., {Schmidtobreick} L., {G{\"a}nsicke} B.~T. 2007(B), MNRAS 374, 1359
\bibitem{schmidtobreicketal12-1}
{Schmidtobreick} L., {Rodr{\'{\i}}guez-Gil} P., {Long} K.~S., {G{\"a}nsicke} B.~T., {Tappert} C., {Torres} M.~A.~P. 2012(A), MNRAS 422, 731
\bibitem{schmidtobreicketal12-2}
{Schmidtobreick} L., {Rodr{\'{\i}}guez-Gil} P., {G{\"a}nsicke} B.~T. 2012(B), Mem.\ Societa Astronomica Italiana 83, 610
\bibitem{schmidtobreicketal03-1}
{Schmidtobreick} L., T., {Tappert} C., {Saviane} I. 2003 MNRAS 342, 145
\bibitem{smithetal98-1}
{Smith} D.~A., {Dhillon} V.~S., {Marsh} T.~R. 1998, MNRAS 296, 465
\bibitem{thorstensen+taylor00-1}
{Thorstensen} J. R., {Taylor} C.J. 2000, MNRAS 312, 629
\bibitem{thorstensenetal91-1}
{Thorstensen} J.~R.,  {Ringwald} F.~A.,  {Wade} R.~A.,  {Schmidt} G.~D.,
  {Norsworthy} J.~E.,  1991, AJ 102, 272
\bibitem{tovmassianetal14-1}
Tovmassian G., {Stephania Hernandez} M., {Gonz{\'a}lez-Buitrago} D., {Zharikov} S., {Garc{\'{\i}}a-D{\'{\i}}az} M.~T., 2014, AJ 147, 68
\bibitem{townsley+gaensicke09-1}
Townsley D.~M. and {G{\"a}nsicke} B.~T., 2009, ApJ 693, 1007
\bibitem{warner99-1}
Warner B., 1999, ASP Conf. Ser. 157, 63
\bibitem{watsonetal06-1}
{Watson} C.~A., {Dhillon} V.~S., {Shahbaz} T., 2006, MNRAS 368, 637
\bibitem{watsonetal07-1}
{Watson} C.~A., {Steeghs} D., {Dhillon} V.~S., {Shahbaz} T., 2007, AN 328, 813
\bibitem{wuetal95-3}
{Wu} K., {Wickramasinghe} D.\ T., {Warner} B., 1995, PASA 12, 60
\bibitem{zangrillietal97-1}
{Zangrilli} L., {Tout} C.~A., {Bianchini} A., 1997, MNRAS 289, 59

\end{thebibliography}
\end{document}